# Assessing the neurocognitive correlates of resting brain entropy

Ze Wang, PhD

Department of Diagnostic Radiology and Nuclear Medicine

University of Maryland School of Medicine

670 W. Baltimore St, Baltimore, MD 20201

ze.wang@som.umaryland.edu

**Abstract**

The human brain exhibits large-scale spontaneous fluctuations that account for most of its total energy metabolism. Independent of any overt function, this immense ongoing activity likely creates or maintains a potential functional brain reserve to facilitate normal brain function. An important property of spontaneous brain activity is the long-range temporal coherence, which can be characterized by resting state fMRI-based brain entropy mapping (BEN), a relatively new method that has gained increasing research interest. The purpose of this study was to leverage the large resting state fMRI and behavioral data publicly available from the human connectome project to address three important but still unknown questions: temporal stability of rsfMRI-derived BEN; the relationship of resting BEN to latent functional reserve; associations of resting BEN to neurocognition. Our results showed that rsfMRI-derived BEN was highly stable across time; resting BEN in the default mode network (DMN) and executive control network (ECN) was related to brain reserve in a negative correlation to education years; and lower DMN/ECN BEN corresponds to higher fluid intelligence and better task performance. These results suggest that resting BEN is a temporally stable brain trait; BEN in DMN/ECN may provide a means to measure the latent functional reserve that bestows better brain functionality and may be enhanced by education.

**Introduction**

The human brain is a dynamic system with large-scale ongoing fluctuations. Understanding these fluctuations is essential to understanding the individual differences of brain function, functional anatomy, and the pathologies associated with neuropsychiatric conditions. Both theoretical models and neuroscience experiments have demonstrated a characteristic self-organized criticality of normal brain activity [1, 2]. A crucial aspect of this criticality is the emergence of long range temporal correlations (LRTC), which have been shown to be fundamental to high-order brain functions such as memory, attention, perception, coordination, etc [3-23]. Loss of temporal coherence may cause inter-neuronal and inter-regional dysconnections. Restoring these dysconnections and the related dysfunctions may require a restoration of LRTC of brain activity. In fact, a recent study has shown that enhancing brain activity coherence improved memory in older people [24]. Over the past decades fMRI, especially resting state fMRI, has been used predominantly to elucidate the potential importance of LRTC by focusing on slow fluctuations in fMRI timeseries and the intrinsic spatial modes that they define, for example, the default mode [25]. Rather than assessing only the slow fluctuations, we have proposed a method [26] to directly map the whole brain LRTC using a nonparametric entropy metric, the Sample Entropy [27, 28]. This metric is based upon the entropy of measured hemodynamic states that considers dependency over time through temporal embedding and long range similarity matching. In other words, this use of entropy reflects the statistical dependencies or order implicit in itinerant dynamics, expressed over extended periods of time. Brain entropy (BEN) mapping results based on resting state fMRI (rsfMRI) have been shown to be unrelated to regional perfusion and other rsfMRI measures in most parts of the brain cortex [29]. Resting BEN is reproducible across time and sensitive to various brain diseases and to focal neuromodulations [26, 30-32]. While these data clearly demonstrate the potential of BEN, a direct measure of LRTC as a unique brain signature for studying brain diseases or normal brain conditions, its neuro-mechanism remains unclear as does how stable this relatively new brain measure is across time.

The purpose of this study was to address the above questions using the large rsfMRI and behavioral data from the human connectome project (HCP) [33]. It has been proposed that the role of resting state brain activity is to facilitate overt brain functions [25, 34-36]. Although it is unclear how such facilitation works, this energy metabolism-costly process (resting state activity accounts for most brain energy metabolism [25, 37]) may actually generate or maintain a brain functionality reserve, or equivalently a collection of various functional brain states. Constantly shuffling among these latent states may act as a priming condition used to respond to upcoming familiar or novel events. Given the aforementioned important role of LRTC in brain activity, we hypothesized that LRTC of resting state brain activity as measured by BEN is related to the latent brain reserve with lower BEN (meaning greater LRTC) indicating a bigger or stronger reserve. Because the latent reserve and resting state activity are both non-specific to any particular cognitive function, they should correlate with general cognitive capabilities and many functionalities. To test the potential role of resting BEN as an index of the latent functional reserve, we will examine the correlations between resting BEN and education and fluid intelligence that are associated with general functional capability and intelligence. Education is a major indicator of cognitive reserve [38], a concept used to explain the individual difference of brain resilience to neuropathology in Alzheimer's Disease. For young healthy individuals, education is known to be strongly correlated with general intelligence [39], suggesting it as a sensitive index for assessing the latent brain functionality reserve. Fluid intelligence is the capability for solving newly encountered problems for which learned and specialized skills provide little benefit [40]. Given the fact that fluid intelligence and the latent functional reserve are two general properties of the brain, it is reasonable to expect that they are correlated. For the same reason, we hypothesized that the latent reserve as measured by resting BEN is correlated to various functional task performances showing non-specificity to a particular functional domain.

Resting BEN has been shown to be replicable across two different acquisition times [26] but the length of rsfMRI time series is short (<130), making it impossible to assess the temporal variations of BEN. By contrast, the HCP rsfMRI data has 1200 timepoints, providing sufficient data for assessing the dynamic information of BEN. We hypothesized that BEN is a stable brain trait presenting small variations across time.

In addition to the above questions, we re-examined the age effects of BEN that have been reported before using small samples [29, 41].

**Materials and Methods**

rsfMRI data, demographic data, and behavior data from 860 healthy young subjects (age 22-37 yrs, male/female=398/464) were downloaded from HCP. Each subject had four resting scans acquired with the same multi-band sequence[42] but the readout directions differed: readout was from left to right (LR) for the 1$^{st}$ and 3$^{rd}$ scans and right to left (RL) for the other two scans. The pre-processed rsfMRI data in the Montreal Neurological Institute (MNI) brain atlas space were downloaded from HCP and were smoothed with a Gaussian filter with full-width-at-half-maximum = 6mm to suppress the residual inter-subject brain structural difference after brain normalization and artifacts in rsfMRI data introduced by brain normalization. BEN mapping was performed with BEN mapping toolbox (BENtbx) using the default settings [26]. To cope with the huge computation required to calculate BEN for the 4x860 long rsfMRI scans (each with 1000 timepoints), we implemented the BEN mapping algorithm in C++ using CUDA (the parallel computing programming platform created by Nvidia Inc). Four graphic processing unit (GPU) video cards were used to further accelerate the process. Entropy value was calculated using the approximate entropy formula, the Sample Entropy, which is the "logarithmic likelihood" that a small section (within a window of a length 'm') of the data "matches" with other sections will still "match" the others if the section window length increases by 1 (see Fig. 1B). "Match" is defined by a threshold of r times standard deviation of the entire time series. In this study, the window length was set to be three and the cut off threshold was set to 0.6 (Wang et al., 2014).

Mean BEN maps of the first LR and the first RL scans and the second LR and the second RL scans were calculated for the following analyses. Age, sex, and education associations of resting BEN were assessed with simple regression using SPM (https://www.fil.ion.ucl.ac.uk/spm/). Associations of BEN to fluid intelligence (measured by the Penn Matrix Test [43]), and functional task performance were similarly examined but with age and sex included as nuisance covariates. Task performance was measured by the accuracy of button selection during the on-magnet fMRI-based working memory, language, and relational tasks[44]. The voxelwise significance threshold was defined by p<0.05. Multiple comparison correction was performed with the family wise error theory [45].

To assess the temporal stability of BEN, we implemented a sliding-window based BEN mapping algorithm. Fig. 1 provides a schematic view of this new algorithm. Similar to the current static BEN mapping, entropy calculation in dynamic BEN mapping is performed at each voxel separately. A time window with a length of L timepoints (L=8 in Fig. 1) sliding from the beginning to the end of the original time series is used to extract a set of temporally overlapped sub-series with one sub-series at each sliding position (Fig. 1A). For each sub-series, the regular SampEn calculation (Fig. 1B, 1C) is applied to get the entropy value at the corresponding sliding window position. "m" in Fig. 1B indicates the window length for SampEn calculation. Fig. 1B.1 illustrates the process of finding the total number of matches among all possible embedding vectors (the temporal signal segments extracted by the smaller sliding window of a length of m). Fig. 1B.2 is a repetition of Fig. 1B.1 but with the embedding vector length increased by 1. And the final SampEn value becomes the natural logarithm of the ratio between the total number of matches of window length m and window length m+1 (Fig. 1C).

The length of the entire time series was 12000. Because BEN mapping using rsfMRI data with a length from 120 to 200 has been shown to provide reliable results, we chose 300 as the sliding window length to get reliable transit BEN from each 300 timepoints rsfMRI sub-series. Successive sub-series were gapped by 9 timepoints to reduce the total number of sub-series to reduce the total computation burden. This gap was empirically set to be 9 timepoints so the interval was 9TRs=6.48 sec, which was roughly the same as one hemodynamic response function cycle. Similar to the static BEN mapping mentioned above, we implemented the dynamic BEN mapping algorithm in C++ and the CUDA programming environment. GPU computing was used for finding the number of matched vectors for many voxels simultaneously. The number of voxels to be processed simultaneously was determined based on the available computation resource in the GPU card. Four Nvidia 1080Ti GPU cards were used.

After dynamic BEN mapping, each subject had a BEN image series. The mean, standard deviation (STD), and the coefficient of variance (CV) of this BEN image series were calculated. For each one of them, the average across the first LR and RL, and the second LR and RL scans were calculated. Similar statistical analyses as mentioned above were performed to assess the potential associations of these maps to age, sex, and cognition.

**Results**

The GPU-based implementation of BENtbx was 10-fold faster than the original version. But it still took roughly 10 days to calculate the static BEN maps (using all 1200 timepoints) for all 1023 subjects (only 862 had all 4 rsfMRI scans). The dynamic BEN mapping took about 40 days. Fig. 2 shows the mean BEN maps (2A-2D), mean STD maps (2E, 2F), mean CV maps (2G, 2H) of the two sessions (each session containing a LR and a RL scan) of all 862 subjects. The mean BEN maps from the static BEN mapping (shown in Fig. 2A and 2B) were very similar to those from the dynamic BEN maps (Fig. 2C and 2D), although the intensity differed due to the significant difference of the time series length (1200 for the static BEN mapping vs 300 for the dynamic BEN mapping). Gray matter (GM) showed lower BEN than white matter (WM), and regions in the default mode networks (DMN) had lower BEN than the rest of the brain; both findings are consistent with our previous study. Dynamic BEN showed inhomogeneous fluctuations across the brain with higher fluctuations in WM, visual cortex, and motor cortex (Fig. 2E, 2F). In relative to the mean BEN, dynamic BEN showed very high temporal stability as measured by the CV (<0.032 in the whole brain, Fig. 2G, 2H. Data with CV <1 is often considered low variation).

We then assessed the effects of age and sex on BEN and its variations. We also examined the associations between cognition and BEN as well as its variations. The results from the BEN maps calculated from the entire 1200 time points and then averaged across the LR and RL scans were highly similar to those from the mean BEN maps of the dynamic BEN image series. Therefore, the results shown below were based on the static BEN mapping results. Also because the results based on the mean of the first LR and the first RL rsfMRI scans or the mean of the second LR and the second RL scans were very similar, we chose to show the results based on the mean BEN of the first LR and RL scans only.

Fig. 3 shows the association maps of BEN to age, sex, education years, and fluid intelligence. Resting BEN was significantly correlated with age (Fig. 3A) in the prefrontal executive control network (ECN, consisting of the lateral prefrontal cortex, the posterior parietal cortex, the frontal eye fields, and part of the dorso-medial prefrontal cortex) and the frontal-temporal-parietal DMN. Women showed higher BEN in visual cortex, motor area, and some part of precuneus (Fig. 3B) than men. Longer education years were associated with decreased BEN in ECN and DMN (Fig. 3C). In Fig 3D, higher fluid intelligence was associated with lower BEN in part of ECN and DMN.

Fig 4 shows the associations of resting BEN to functional task performance. BEN in DMN and part of ECN was negatively correlated with better performance during performing working memory (Fig. 4A), language (Fig. 4B), and relational tasks (Fig. 4C). Temporal STD of BEN showed no significant age and sex effects and no significant correlations to education years, fluid intelligence, and task performance.

## Discussion

In this study, we assessed the long-range temporal coherence of resting state brain activity using a large data set. Long-range coherence was measured by BEN. A sliding window-based dynamic BEN mapping method was implemented to examine the temporal fluctuations of BEN. Our data showed that BEN was stable across the entire acquisition time with minor temporal variations, and did not show any significant correlation to age, sex, education, and neurocognitive measures. To understand the potential neuro-cognitive mechanism of resting BEN, we assessed the associations of BEN with biological and behavioral measures and found that BEN in DMN and ECN increases with age but decreases with years of education; women had higher BEN than men in the cortical area; BEN in DMN and ECN was negatively correlated with fluid intelligence and task performance for all of the assessed cognitive tasks.

The high temporal stability of resting BEN across many different timepoints is consistent with the high test-retest reproducibility of BEN shown in our previous study [26], further proving BEN to be a reliable brain metric. Our findings of a strong positive correlation between age and BEN in DMN and ECN were consistent with the results reported in [46] and provide additional evidence for the physical law-based brain entropy hypothesis which states that brain entropy tends to increase with time in the normal adult brain due to progressive tissue aging and deteriorations[47-51]. While this unfortunate entropy increase trend may seem avoidable, our data also showed that longer education years were correlated with lower resting BEN, suggesting a plausible way of reducing resting BEN through extended learning. Our previous study showed that beneficial focal stimulations via transcranial magnetic stimulations can reduce local BEN[52]. The education effects on resting BEN revealed in this paper further proved that resting BEN is modifiable, which is of particular interest for future brain disease studies.

The finding of females having higher BEN than males was consistent with [41], which might be due to hormonal effects [53]. The sex difference of BEN in the visual and motor cortex may reflect the sex difference of visual and motion processing previously reported in [54].

Our BEN vs education and cognition association analysis results unanimously highlighted DMN and ECN, which is consistent with the well-known phenomenon that DMN and ECN (also called task positive network [55]) are two major brain circuits that are both active either during task performance [55, 56] or at rest [57, 58]. Different from the previous studies, our results suggest that both DMN and ECN are related to neurocognition through their resting BEN, which is independent of age, sex, and education though all three factors did show significant effects on DMN and ECN BEN. In terms of the long-range temporal coherence, DMN and ECN may actually represent a unified neural circuit underlying the general intelligence and general functionality of the brain.

Resting state brain activity has been postulated to be involved in maintaining and facilitating brain functions such as language, social interaction, and memory [25, 59-61]. Our data directly support those postulations through the negative correlations between resting BEN and the task performance for three different functional tasks. Moreover, our data for the first time linked regional resting activity as measured by BEN to general intelligence as reflected by fluid intelligence and education years.

Regarding the correlation between resting brain activity and task activations, several fMRI studies have reported that brain activation during functional task performance can be predicted reliably by the resting state fMRI based on regional inter-voxel correlations[62, 63], the amplitude of the low frequency fluctuations[64], and inter-regional functional connectivity[65-69]. Our study differs from these by assessing the brain-behavior associations rather than a brain vs brain relationship. Several groups have reported the correlations between resting state functional connectivity and task behavior or cognition[63, 70-72].

## Conclusion

In conclusion, the long rsfMRI time series from a large cohort of healthy subjects in the HCP proved BEN is a temporally stable brain activity measure. Our data suggest BEN in DMN/ECN can be used as a measure of the potential functional reserve that can be improved by education and may result in better brain function.

## Acknowledgements

The research effort involved in this study was supported by NIH/NIA R01 AG060054 and by the University of Maryland, Baltimore, Institute for Clinical & Translational Research (ICTR). Both imaging and behavior data were provided by the Human Connectome Project, WU-Minn Consortium (Principal Investigators: David Van Essen and Kamil Ugurbil; 1U54MH091657) funded by the 16 NIH Institutes and Centers that support the NIH Blueprint for Neuroscience Research; and by the McDonnell Center for Systems Neuroscience at Washington University in St. Louis. The authors thank the Human Connectome Project for open access to its data and thank Brigitte Pocta for editing the manuscript. Part of the literature review in Introduction was generously provided by Prof Karl J. Friston.

reports, 2019. **9**(1): p. 1-16.

**Legends**

Fig. 1. Resting fBEN was A) positively correlated with age, B) higher in females, C) negatively correlated with education and D) negatively correlated with fluid intelligence C). n=860. P<0.05 (FWE corrected).

Fig. 2. A scheme of the sliding window-based dynamic entropy calculation. A) A large time window is used to extract a sub-time series at N successive timepoints (N=8 here) from the original time series. The green box indicates the window slid to the n-th timepoint. B) The standard sample entropy formula is used to calculate entropy for the sub-time series extracted from A. B.1 and B.2 illustrate the embedding vector matching process for the embedding window length of m and m+1, respectively. The boxes in different color indicate the locations of the embedding vectors in the input time series—the sub-series from A).

Fig. 3. The age, sex, and education effects on resting BEN as well as the associations of resting BEN with fluid intelligence. n=862. p<0.05 (FWE corrected). Color bars indicate the t-values of the regression analysis (for A and C) and the two-sample t-tests (B and D).

Fig. 4. Resting fBEN was negatively associated with task performance levels for A) working memory task, B) language task, C) relational task.

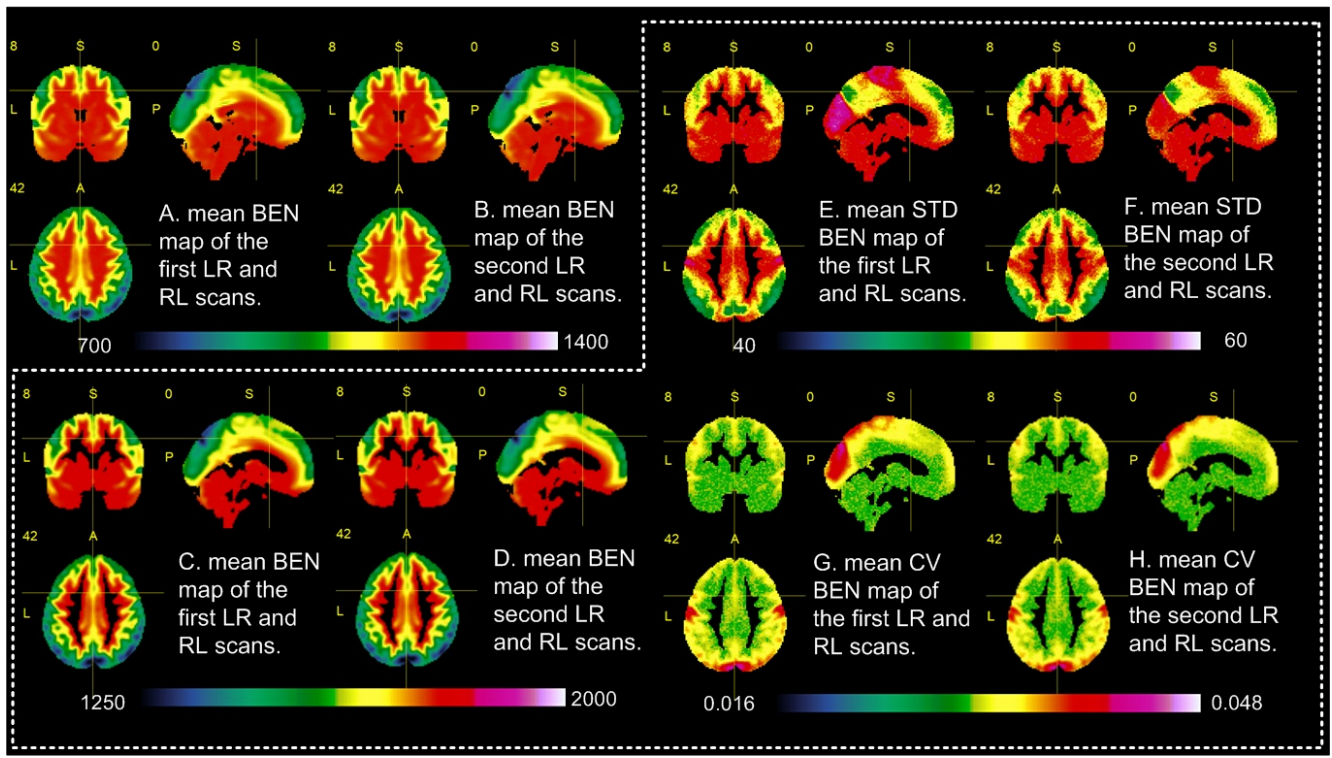

Fig. 1. Resting fBEN was A) positively correlated with age, B) higher in females, C) negatively correlated with education and D) negatively correlated with fluid intelligence C). n=860. P<0.05 (FWE corrected).

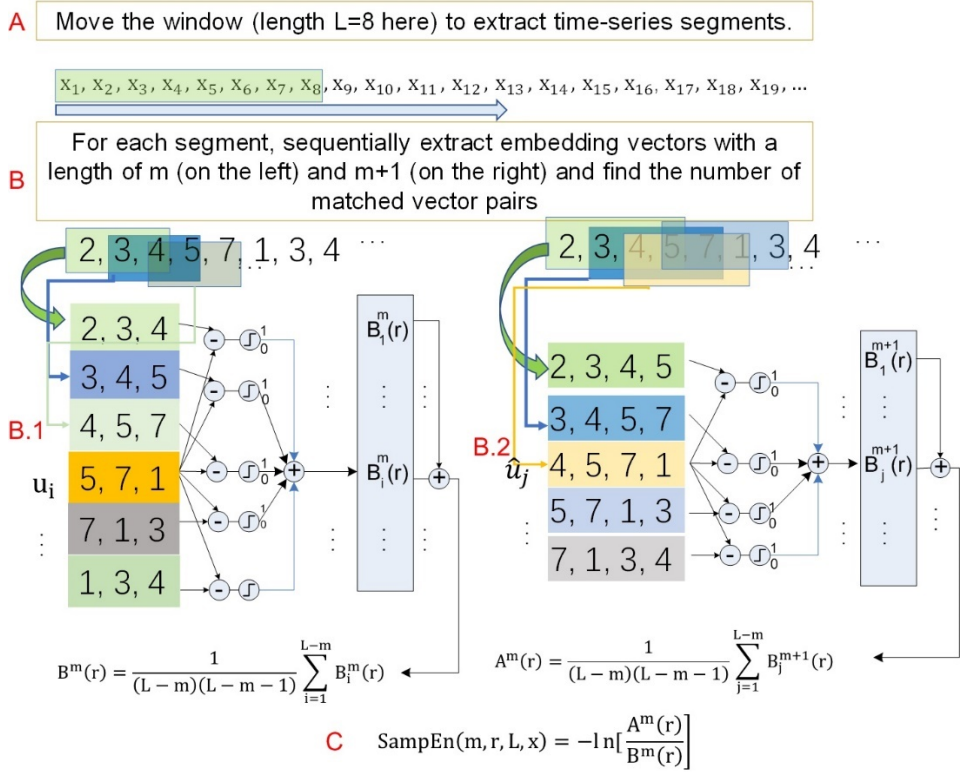

Fig. 2. A scheme of the sliding window-based dynamic entropy calculation. A) A large time window is used to extract a sub-time series at N successive timepoints (N=8 here) from the original time series. The green box indicates the window slid to the n-th timepoint. B) The standard sample entropy formula is used to calculate entropy for the sub-time series extracted from A. B.1 and B.2 illustrate the embedding vector matching process for the embedding window length of m and m+1, respectively. The boxes in different color indicate the locations of the embedding vectors in the input time series—the sub-series from A).

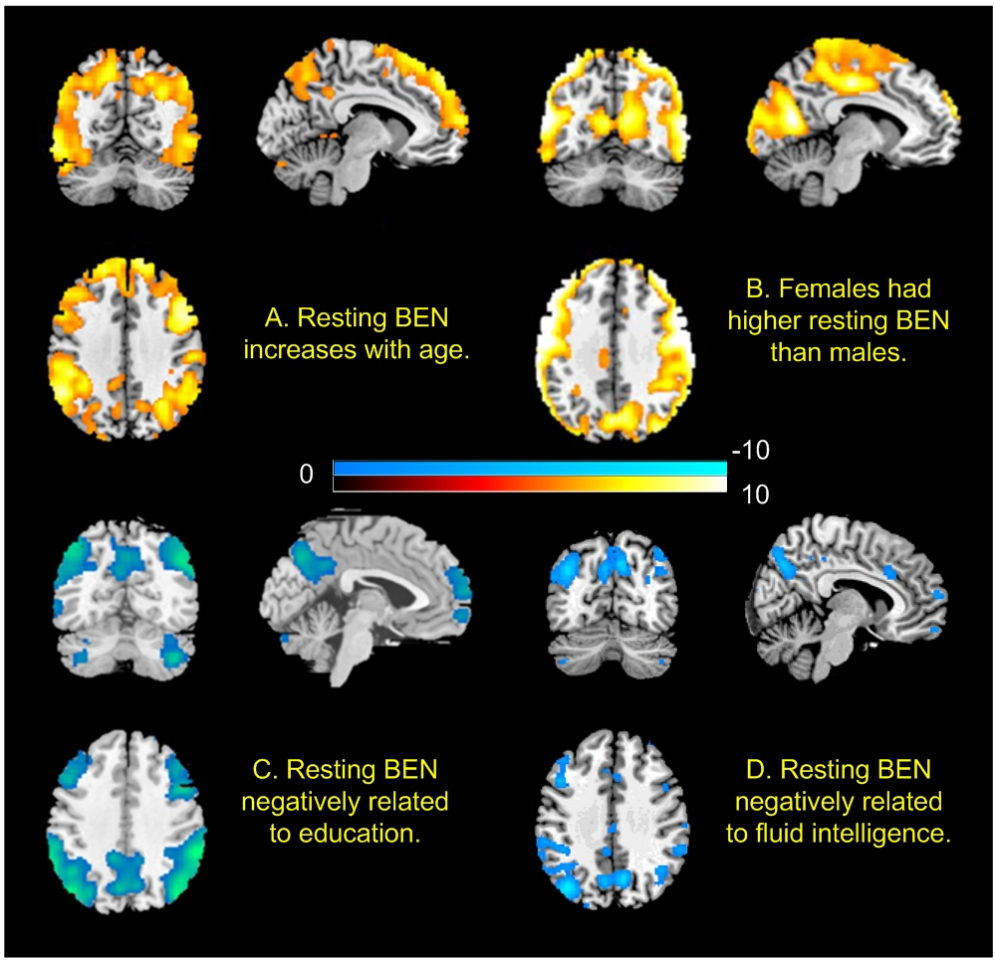

Fig. 3. The age, sex, and education effects on resting BEN as well as the associations of resting BEN with fluid intelligence. n=862. p<0.05 (FWE corrected). Color bars indicate the t-values of the regression analysis (for A and C) and the two-sample t-tests (B and D).

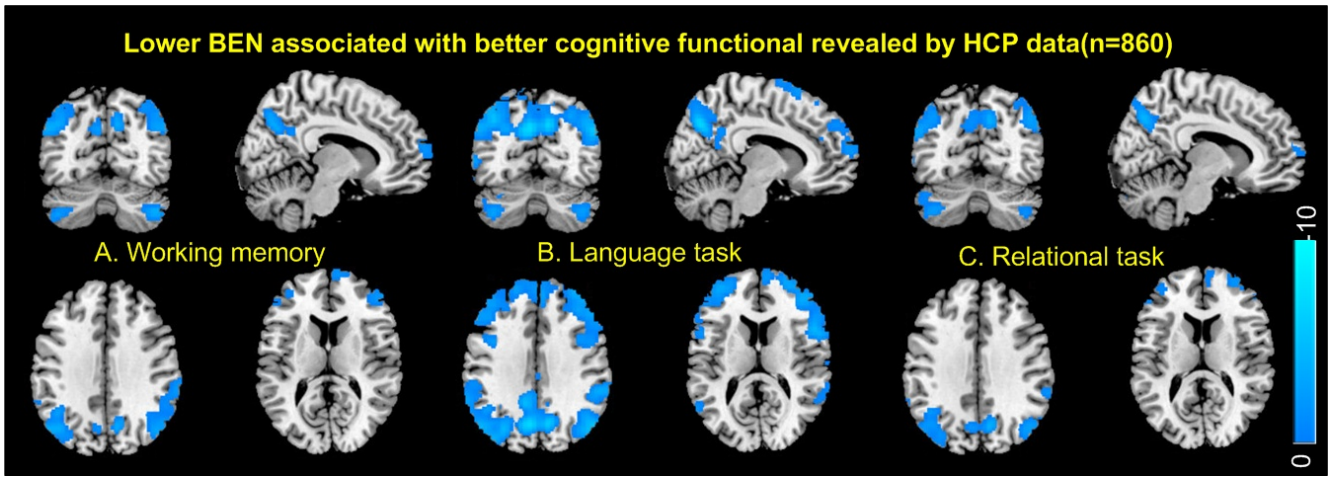

Fig. 4. Resting fBEN was negatively associated with task performance levels for A) working memory task, B) language task, C) relational task.